# Investigation of Chaotic Behavior in Clapp Oscillator

Ivana Vasiljević, Nikola Petrović and Aleksandra Lekić, *Member, IEEE*

***Abstract*—In this paper we investigate the chaotic behavior of the class of oscillators denoted as Clapp oscillators. Clapp oscillator is a simple oscillator containing one transistor and a few reactive elements - inductors and capacitors. This oscilllator is chosen for its design simplicity and a good performance. Oscillator with chaotic behavior can be used to construct chaotic radar. For that matter, in this paper is investigated approach for construction of the chaotic Clapp oscillator, which can be further verified experimentally using microstrip technology.**

## I. INTRODUCTION

THE idea for construction of the chaotic radar grew special attention after discovery of the self-synchronization of the chaotic systems [1]. This discovery brought in the focus research topics related to application of the chaotic signals in communication systems, especially for the construction of the chaotic synchronized systems [2, 3]. For that matter, the circuits which have possibility of chaotic operation were special interesting, i.e. oscillators and their chaotic character.

One of the pioneers in the area of investigation chaotic behavior is Leon Chua. Chua studied chaotic behavior of the oscillator named later after him, Chua's circuit [4, 5]. Attractors obtained during chaotic operation in Chua's circuit have been analyzed in detail in [6, 7] using the theory of nonlinear systems and a chaos theory [8]. Besides Chua's circuit, there are a number of well-known circuits which have oscillating chaotic character. Two types of oscillators are known as Hartley's [9, 10] and Colpitts oscillator [2, 3, 11-16]. Hartley's and Colpitts oscillators contain one transistor and a few bias resistors, capacitors and inductors, which make them simple enough for both construction and analysis.

In this paper we have chosen Clapp oscillator [17] depicted in Fig. 1, which is very similar to the Colpitts oscillator and the only difference is the existence of the capacitor $C_3$. Meaning that the Clapp oscillator is a series tuned version of the Colpitts oscillator. Clapp oscillator is designed and analyzed such that it can produce chaotic operation and thus, it can be incorporated into chaotic radar [2, 3, 18]. The operation of the Clapp oscillator will be demonstrated in the similar matter as the improved Colpitts oscillator reported in [15]. Analysis is done on the level of system linearization, calculation of eigenvalues of the linearized system and checking their stability. Finally, the chaotic behavior is determined by the calculation of linearized system eigenvalues comparing to relevant circuit parameters as described in [8, 15]. For the Clapp oscillator, circuit values which guarantee desirable performance will be provided.

The paper is organized as follows. Section II provides derivation of the Clapp circuit equations and oscillating conditions. The simulation diagrams showing chaotic performance are provided for the selected transistor and the other circuit parameters. Section III contains description of the oscillator linearization and numerical simulation used for calculation of the eigenvalues of the linearized circuit. The section IV gives concluding remarks.

Ivana Vasiljević, Nikola Petrović and Aleksandra Lekić are with the School of Electrical Engineering, University of Belgrade, 73 Bulevar kralja Aleksandra, 11020 Belgrade, Serbia (e-mail: ivana9558@gmail.com, p.z.nikola@etf.bg.ac.rs, lekic.aleksandra@etf.bg.ac.rs).

## II. MODEL DERIVATION

For the Clapp oscillator depicted in Fig. 1 following equations for transistor currents using the model for large signal bipolar junction transistor (BJT) can be written:

$$i_C = I_S\left(e^{\eta\frac{v_{BE}}{V_T}}-1\right) = I_S\left(e^{\eta\frac{v_{C1}}{V_T}}-1\right), \quad (1)$$

$$i_E = i_B + i_C = i_B + \beta i_B = (\beta+1)i_B, \quad (2)$$

$$i_B = \frac{i_C}{\beta} = \frac{I_S}{\beta}\left(e^{\eta\frac{v_{C1}}{V_T}}-1\right). \quad (3)$$

In the equilibrium [8], denoted in further text adding *eq* in the index, a base current can be represented as

$$i_{B,eq} = \frac{I_S}{\beta}\left(e^{\eta\frac{v_{C1,eq}}{V_T}}-1\right). \quad (4)$$

Using the previously derived equations (1)-(4) for the

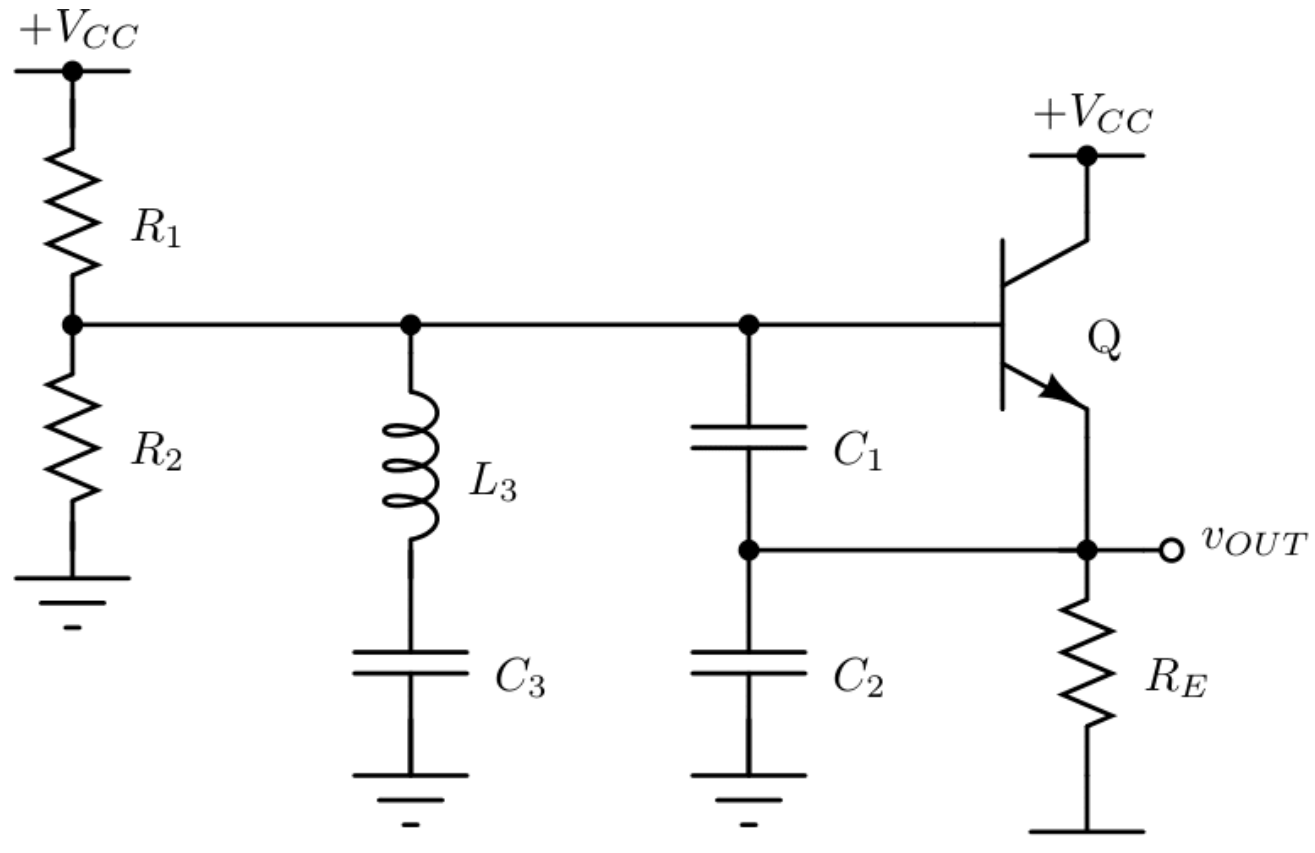

Fig. 1. Clapp oscillator.

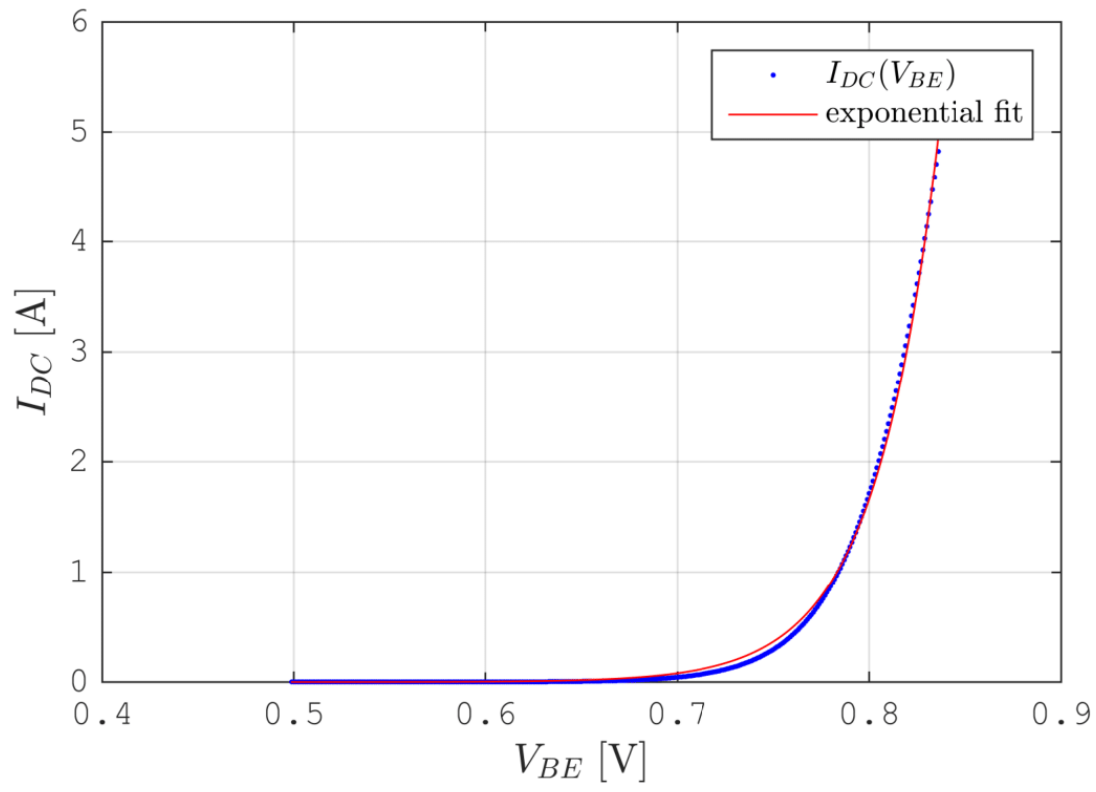

Fig. 2. Dependence $I_{DC}(V_{BE})$ of the transistor BFU730F – dashed blue line and its exponential fit – red line.

circuit depicted in Fig. 1 the following differential equations in the state-space can be written:

$$\frac{dv_{C1}}{dt}=\frac{1}{C_1}\left(-\left(v_{C1}+v_{C2}\right)\left(\frac{1}{R_1}+\frac{1}{R_2}\right)-i_{L3}-i_B+\frac{V_{CC}}{R_1}\right), \quad (5)$$

$$\frac{dv_{C2}}{dt}=\frac{1}{C_2}\left(-\left(v_{C1}+v_{C2}\right)\left(\frac{1}{R_1}+\frac{1}{R_2}\right)-\frac{v_{C2}}{R_E}-i_{L3}+\beta i_B+\frac{V_{CC}}{R_1}\right), \quad (6)$$

$$\frac{dv_{C3}}{dt}=\frac{1}{C_3}i_{L3}, \quad (7)$$

$$\frac{di_{L3}}{dt}=\frac{1}{L_3}\left(v_{C1}+v_{C2}-v_{C3}\right). \quad (8)$$

Capacitor voltages and inductor current represent state variables of the system. Vector of the state-space variables is then $\mathbf{p}=\left[v_{C1}\, v_{C2}\, v_{C3}\, i_{L3}\right]^T$. This system has a single equilibrium point $\mathbf{p}_{eq}=\left[v_{C1,eq}\, v_{C2,eq}\, v_{C3,eq}\, i_{L3,eq}\right]^T$, which is obtained by setting the right hand side of system (represented as a set of equations (5)-(8)) to zero, $\dot{\mathbf{p}}=0$. Equilibrium is given in the following equations:

$$v_{C1,eq}=\frac{\frac{V_{CC}}{R_1}-i_{B,eq}}{\frac{1}{R_1}+\frac{1}{R_2}}-(1+\beta)R_E i_{B,eq}, \quad (9)$$

$$v_{C_2,eq}=(1+\beta)R_E i_{B,eq}, \quad (10)$$

$$v_{C3,eq}=\frac{\frac{V_{CC}}{R_1}-i_{B,eq}}{\frac{1}{R_1}+\frac{1}{R_2}}, \quad (11)$$

$$i_{L_3,eq}=0. \quad (12)$$

Equations (9)-(12) with the addition of equation (4) provide a complete solution for the equilibrium. However, this is the set of transcendent equations and in this work it will be solved numerically in MATLAB.

### A. Estimation of the circuit values

For the purpose of the realization of the Clapp oscillator, it is necessary to select a circuit transistor. For that matter, the transistor BFU730F is chosen. This transistor operates with the best performance at the frequency $f=5.8$ GHz. In order to characterize oscillator, simulations of the dependence $I_{DC}(V_{BE})$ are performed, where $I_{DC}$ is collector current and $V_{BE}$ is the voltage between base and emitter. Mentioned simulations give us parameters of the best approximation of the $I_{DC}(V_{BE})$ dependence in exponential form as depicted in Fig. 2. Obtained parameters for the BJT large signal model, as in equation (1), are $I_S=47.1$ pA and $\eta=0.7894$.

### B. Clapp oscillator circuit parameters

The Clapp oscillator is constructed using transistor BFU730F. For the supply voltage of $V_{CC}=12$ V, biasing resistors $R_1=5$ kΩ and $R_2=7$ kΩ, and emitter resistance $R_E=500$ Ω, capacitors and inductor in the circuit are chosen to operate at the resonant frequency $f=5.8$ GHz periodically. Clapp oscillator from Fig. 1 operates in chaos when: $C_1=C_2=2$ pF, $C_3=0.1$ pF and $L_3=0.753$ nH. The operation of the oscillator is simulated on the room temperature, and thus the effect of the temperature change is not considered. Chaotic behavior in the Clapp oscillator with the selected circuit parameters is visible from the time diagrams depicted in Fig. 3 and from diagrams of the selected state variables with the respect to another state variable depicted in Fig. 4.

Although the diagrams from the Fig. 3 and Fig. 4 show very large voltage on the capacitor $C_3$, the sum of the voltages $v_{C3}+v_{L3}$ is less than the supply voltage. Also, large values of capacitors voltages are due to the fact that transistor model is not limited (i.e. transistor is always working in forward-active mode and cannot enter saturation mode). However, from the time diagram for the voltage $v_{C3}$ we can clearly see

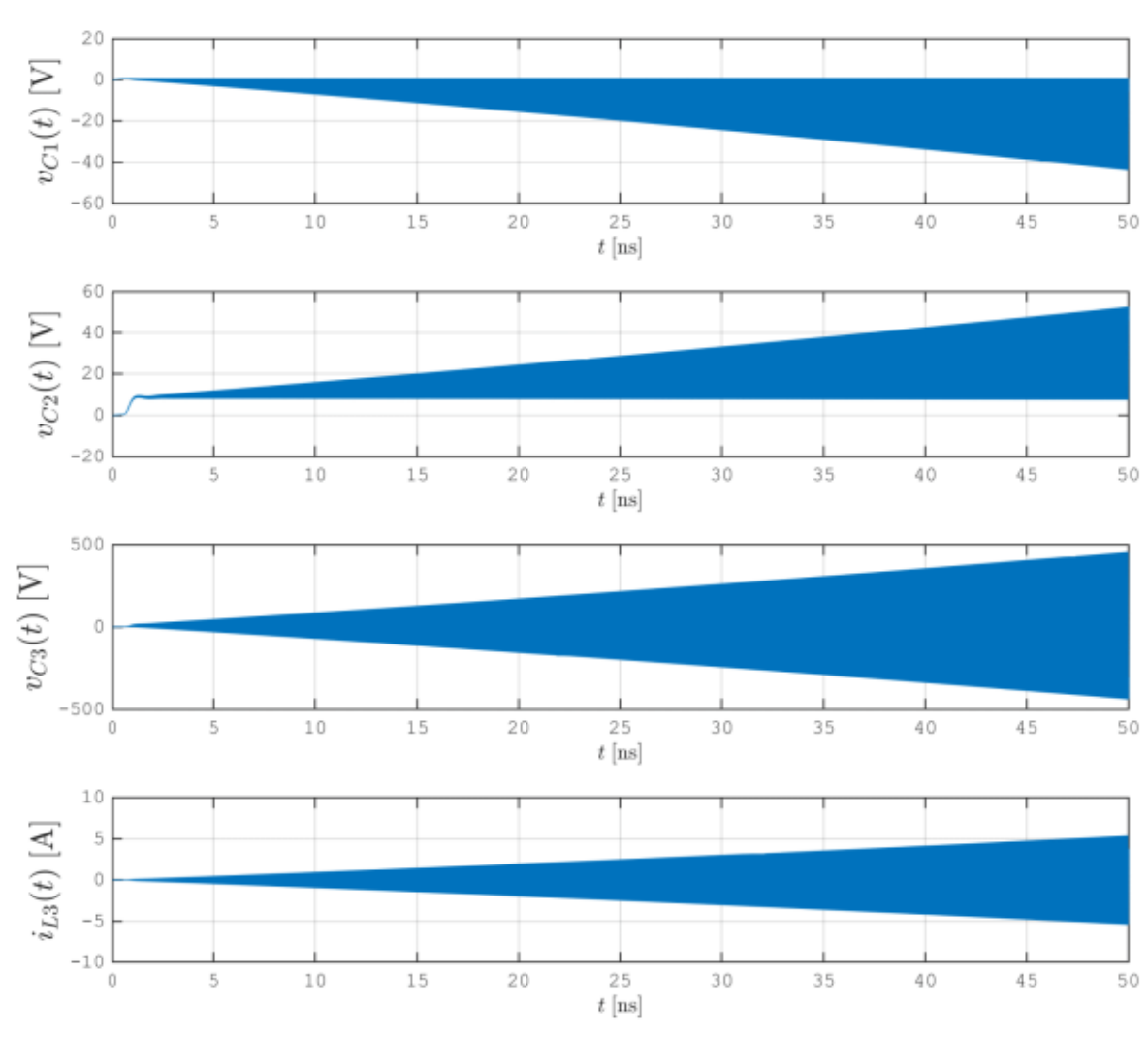

Fig. 3. Time diagrams of the Clapp oscillator operating in chaos.

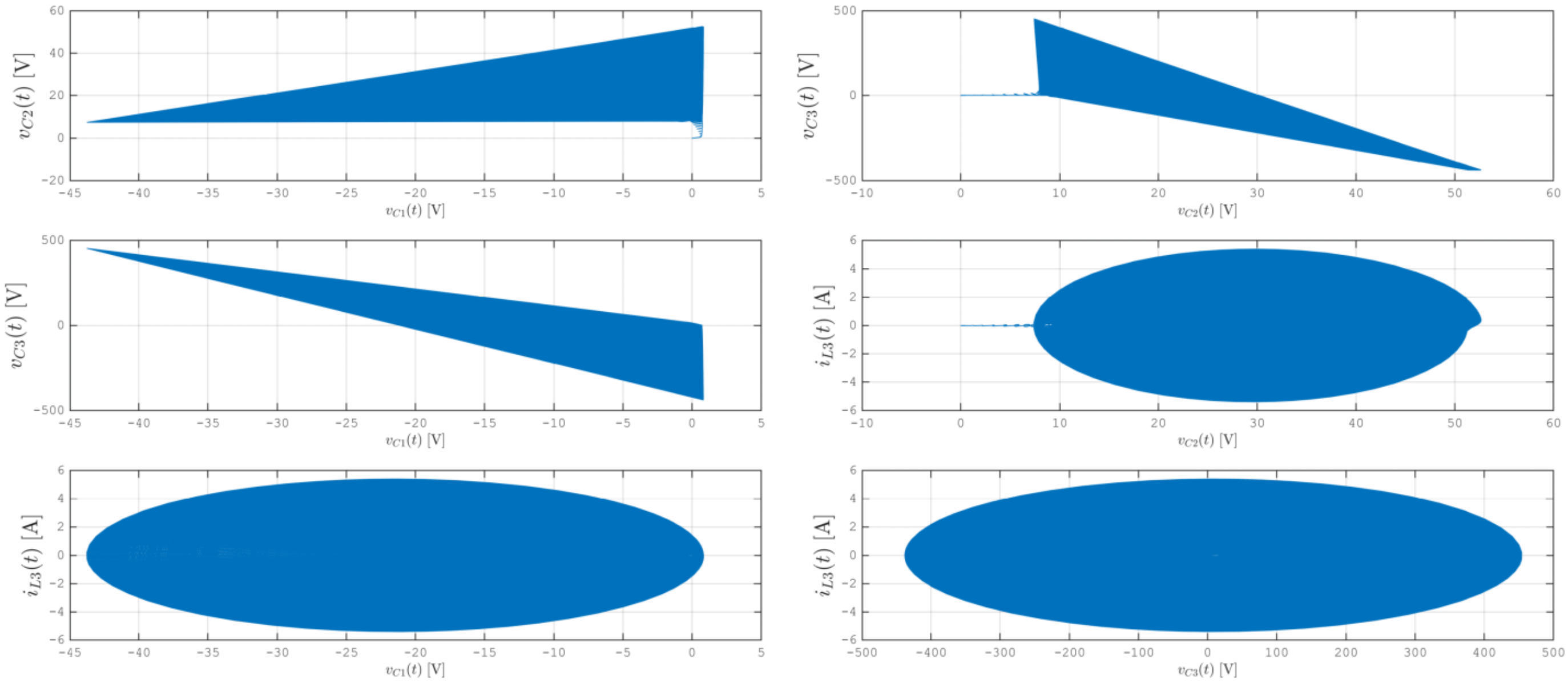

Fig. 4. Dependence one of the state variables with respect to another in the Clapp oscillator operating in chaos.

the divergence of the overall system. Meaning that the oscillator operates in chaos.

## III. Linearization with Jacobian Matrix

Equations of the system in the state-space form given with (5)-(8) represent a set of nonlinear differential equations. A common way to analyze such system is to represent it as a sum of linear and nonlinear part [8]. Linearization of the system is done by deriving system's Jacobian matrix. For easier definition of the Jacobian matrix, the following equations have been introduced:

$$f_1(p)=\frac{dv_{C1}}{dt},\quad f_2(p)=\frac{dv_{C2}}{dt},$$
$$f_3(p)=\frac{dv_{C3}}{dt},\quad f_4(p)=\frac{di_{L3}}{dt}. \tag{13}$$

Jacobian matrix is given as:

$$\mathbf{J}=\left.\begin{bmatrix} \frac{\partial f_1}{\partial v_{C1}} & \frac{\partial f_1}{\partial v_{C2}} & \frac{\partial f_1}{\partial v_{C3}} & \frac{\partial f_1}{\partial i_{L3}} \\ \frac{\partial f_2}{\partial v_{C1}} & \frac{\partial f_2}{\partial v_{C2}} & \frac{\partial f_2}{\partial v_{C3}} & \frac{\partial f_2}{\partial i_{L3}} \\ \frac{\partial f_3}{\partial v_{C_1}} & \frac{\partial f_3}{\partial v_{C_2}} & \frac{\partial f_3}{\partial v_{C_3}} & \frac{\partial f_3}{\partial i_{L_3}} \\ \frac{\partial f_4}{\partial v_{C_1}} & \frac{\partial f_4}{\partial v_{C_2}} & \frac{\partial f_4}{\partial v_{C_3}} & \frac{\partial f_4}{\partial i_{L_3}} \end{bmatrix}\right|_{v_{C1,eq},v_{C2,eq},v_{C3,eq},i_{L3,eq}} \tag{14}$$

where $\mathbf{p}_{eq}=\left[v_{C1,eq}\; v_{C2,eq}\; v_{C3,eq}\; i_{L3,eq}\right]^T$ presents an equilibrium point given with equations (9)-(12).

The system can be described by the following equation:

$$\dot{\mathbf{p}}=\mathbf{J}\cdot\mathbf{p}+\mathbf{h}(\mathbf{p}) \tag{15}$$

where $h(\mathbf{p})$ is a nonlinearity.

Using ordinary differential equations (5)-(8) and equilibrium (9)-(12), Jacobian and nonlinearity are:

$$\mathbf{J}=\begin{bmatrix} -\frac{1}{C_1}\left(\left(\frac{1}{R_1}+\frac{1}{R_2}\right)+\frac{\eta I_S}{\beta V_T}e^{\eta\frac{v_{C1,eq}}{V_T}}\right) & -\frac{1}{C_1}\left(\frac{1}{R_1}+\frac{1}{R_2}\right) & 0 & -\frac{1}{C_1} \\ -\frac{1}{C_2}\left(\left(\frac{1}{R_1}+\frac{1}{R_2}\right)-\frac{\eta I_S}{V_T}e^{\eta\frac{v_{C1,eq}}{V_T}}\right) & -\frac{1}{C_2}\left(\frac{1}{R_1}+\frac{1}{R_2}+\frac{1}{R_E}\right) & 0 & -\frac{1}{C_2} \\ 0 & 0 & 0 & \frac{1}{C_3} \\ \frac{1}{L_3} & \frac{1}{L_3} & -\frac{1}{L_3} & 0 \end{bmatrix}, \tag{16}$$

and

$$\mathbf{h}(\mathbf{p})=\begin{bmatrix} \frac{1}{C_1}\left(-\frac{I_S}{\beta}\left(e^{\eta\frac{v_{C1}}{V_T}}-1\right)+\frac{\eta I_S}{\beta V_T}e^{\mu\frac{v_{C1,eq}}{V_T}}v_{C1}+\frac{V_{CC}}{R_1}\right) \\ \frac{1}{C_2}\left(I_S\left(e^{\eta\frac{v_{C1}}{V_T}}-1\right)-\frac{\eta I_S}{V_T}e^{\eta\frac{v_{C1,eq}}{V_T}}v_{C1}+\frac{V_{CC}}{R_1}\right) \\ 0 \\ 0 \end{bmatrix}. \tag{17}$$

Clapp oscillator for the chosen values has clearly unstable eigenvalues, which is determined by finding eigenvalues of the Jacobian given in equation (16). The two eigenvalues are 4030371149.1 and two have value -5605928916.24.

### A. *Estimation of the operation at the edge of chaos*

In order to determine the values of the resistor $R_E$ for which the oscillator operates in chaos, a linearized model around system's equilibrium is simulated. From the diagram depicted in Fig. 5, which shows dependence of the maximum eigenvalues of the linearized system, equation (16), with respect to the resistance $R_E$. It is shown that for the resistance $R_E \geq 18.925\Omega$ the behavior of the system is clearly unstable, because its larges eigenvalue has positive real part. Such analysis represents simplified calculation of the first Lyapunov exponent [8]. The calculated eigenvalues are determined for the linearized system and they represent rough estimation for the chaos boundary. Thus, the lower estimation of the chaos boundary provides guarantee that the system for some resistance value will definitely enter chaos, but system's nonlinearity can make this boundary even lower.

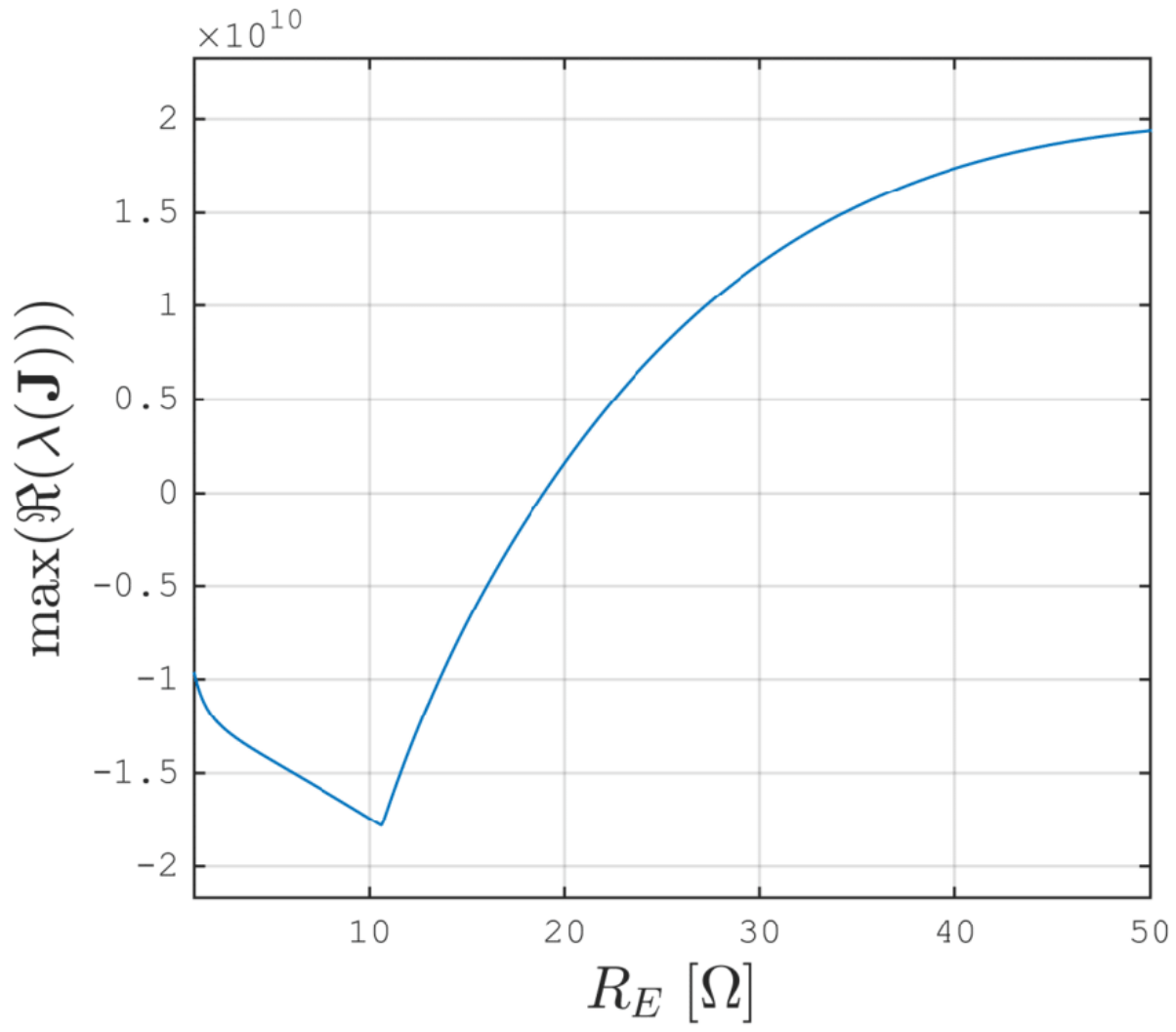

Fig. 5. Dependance of the Jacobian's maximum eigenvalue on resistance $R_E$.

## IV. Conclusion

In this paper we have proposed analysis of the Clapp oscillator operating in the chaos. Constructed Clapp oscillator contains transistor BFU730F. Circuit components are chosen to guarantee chaotic behavior with components whose parameter values are commonly used in microstrip. In the future work the designed oscillator will be implemented and the simulation results will be compared with experimental.

## Acknowledgment

The work has been supported through the Republic of Serbia Ministry of Science project TR33020.